\documentclass[12pt]{article}
\usepackage{psfig,epsf}
\usepackage{a4wide,graphicx}
\usepackage{cite}

\begin{document}

\title{Low lying $S=-1$ excited baryons and chiral symmetry}

\author{
E. Oset \\ 
{\small Departamento de F\'{\i}sica Te\'orica and IFIC,
Centro Mixto Universidad de Valencia-CSIC,} \\ 
{\small Institutos de
Investigaci\'on de Paterna, Aptd. 22085, 46071 Valencia, Spain}\\ 
~\\ 
A. Ramos \\ 
{\small Departament d'Estructura i Constituents de la Mat\`eria,
Universitat de Barcelona,} \\ 
{\small Diagonal 647, 08028 Barcelona, Spain}\\
~\\
C. Bennhold \\
{\small Center for Nuclear Studies and Department of Physics,} \\
{\small The George Washington University, Washington DC 20052,  USA}
}

\date{\today}

\maketitle
\begin{abstract}
The s-wave meson-baryon interaction in the $S = -1$ sector is studied by
means of coupled-channels, using the lowest-order
chiral Lagrangian and the N/D method to implement unitarity. The loops are
regularized using dimensional renormalization. 
In addition to the previously studied $\Lambda (1405)$, employing this chiral 
approach leads to the dynamical generation of two more s-wave hyperon resonances, the
$\Lambda(1670)$ and $\Sigma(1620)$ states.
We make comparisons with
experimental data and look for poles in the complex plane obtaining the
couplings of the resonances to the different final states. This allows us to 
identify the $\Lambda (1405)$ and the $\Lambda(1670)$ resonances with $\bar{K}N$
and $K\Xi$ quasibound states, respectively.

\end{abstract}

\section{Introduction}

The low-energy $K^-N$ scattering and transition to coupled channels is one of
the cases of successful application of chiral dynamics in the baryon
sector.
%\cite{Ga85}. 
The studies of \cite{Kai95} and \cite{Kai97} showed that one could
obtain an excellent description of the low-energy data starting from chiral
Lagrangians and using the multichannel Lippman-Schwinger equation to
account for multiple scattering and unitarity in coupled channels.  By including
all open channels above threshold and fitting a few chiral parameters of
the second-order Lagrangian one could obtain a good agreement with the data at 
low energies.  This line of work was continued in \cite{angels}, where all coupled
channels were included that could be arranged
from the octet of pseudoscalar Goldstone bosons and the baryon ground state octet.
 In Ref.\cite{angels} it was demonstrated that
using the Bethe-Salpeter equation (BSE) with coupled channels and using the 
lowest-order chiral Lagrangians, together with one cut off to regularize the
intermediate meson-baryon loops, a good description of all
low-energy data was obtained.  One of the novel features with respect to other approaches
using the BSE is that the lowest-order meson-baryon amplitudes, playing the role
of a potential, could be factorized on shell in the BSE, and thus the set of
coupled-channels integral equations became a simple set of algebraic equations,
thus technically simplifying the problem.  The justification of this procedure
is seen in a more general way in the treatment of meson-meson interactions using
chiral Lagrangians and the N/D method in \cite{nsd}.  One uses dispersion
relations and shows that neglecting the effects of the left-hand singularity
(also shown to be small there) one needs only the on-shell scattering matrix
from the lowest-order Lagrangian, and  the eventual effects of higher-order 
Lagrangians are accounted for in terms of subtractions in the dispersion
integrals. 
The N/D method has also been recently applied to study pion-nucleon
dynamics in Ref.~\cite{om00}.

    The work of Ref.\cite{angels} was reanalyzed recently \cite{joseulf} 
from the point of view of the N/D method and dispersion relations, leading
formally to the same algebraic equations found in \cite{angels}. There are also
technical novelties in the regularization of the loop function, which is done
using dimensional regularization in Ref.\cite{joseulf}, while it was regularized
with a cut off in Ref.\cite{angels}. 

   One of the common findings shared by all the theoretical approaches is
the dynamical generation of the  $\Lambda(1405)$ resonance which appears with
the right width, and at the correct position, with the choice of a cut off of
natural size. This natural generation from the interaction of the meson-baryon
system with the lowest-order Lagrangian allows us to identify that state as a
quasibound meson-baryon state. This would explain why ordinary quark models have
had so many problems explaining this resonance \cite{quarks}.

   In ordinary quark models the $\Lambda(1405)$ resonance would mostly be a
SU(3) singlet of $J^P=1/2^-$ and there would be an associated octet of
s-wave  excited 
$J^P=1/2^-$ baryons that would include the N*(1535), the  $\Lambda(1670)$,
the $\Sigma(1620)$ and a $\Xi^*$ state.  In the chiral approach one would also 
expect the appearance of such a nonet of resonances. In fact, it appears  
naturally in the approach of Ref.\cite{angels},
with a degenerate octet, when setting all the masses of the octet of stable  
baryons equal on one
side and the masses of the octet of pseudoscalar mesons equal on the other side. 
 Yet, to obtain this 
result it is essential that the coupled 
channels do not omit any of the channels that can be constructed from the
octet of pseudoscalar mesons and the octet of stable baryons.

The lowest-order Lagrangian involving the octet of pseudoscalar mesons and 
the $1/2^+$ baryons is given in 
\cite{Pi95,Eck95,Be95,Mei93}.
% and one has
%
%\begin{eqnarray}
%L_1^{(B)} &=& < \bar{B} i \gamma^{\mu} \nabla_{\mu} B> - M_B <\bar{B} B>
%+
%\nonumber \\
%& & \frac{1}{2} D <\bar{B} \gamma^{\mu} \gamma_5 \left\{ u_{\mu}, B
%\right\} >
%+ \frac{1}{2} F <\bar{B} \gamma^{\mu} \gamma_5 [u_{\mu}, B]>
%\label{eq:chiral}
%\end{eqnarray}
%where the symbol $< \, >$ denotes the trace of SU(3) matrices and
%\begin{equation}
%\begin{array}{l}
%\nabla_{\mu} B = \partial_{\mu} B + [\Gamma_{\mu}, B] \\
%\Gamma_{\mu} = \frac{1}{2} (u^+ \partial_{\mu} u + u \partial_{\mu}
%u^+) \\
%U = u^2 = {\rm exp} (i \sqrt{2} \Phi / f) \\
%u_{\mu} = i u ^+ \partial_{\mu} U u^+
%\end{array}
%\label{eq:u}
%\end{equation}
%
%The SU(3) matrices for the mesons and the baryons are the following
%
%\begin{equation}
%\Phi =
%\left(
%\begin{array}{ccc}
%\frac{1}{\sqrt{2}} \pi^0 + \frac{1}{\sqrt{6}} \eta & \pi^+ & K^+ \\
%\pi^- & - \frac{1}{\sqrt{2}} \pi^0 + \frac{1}{\sqrt{6}} \eta & K^0 \\
%K^- & \bar{K}^0 & - \frac{2}{\sqrt{6}} \eta
%\end{array}
%\right)
%\label{eq:mesons}
%\end{equation}
%
%
%\begin{equation}
%B =
%\left(
%\begin{array}{ccc}
%\frac{1}{\sqrt{2}} \Sigma^0 + \frac{1}{\sqrt{6}} \Lambda &
%\Sigma^+ & p \\
%\Sigma^- & - \frac{1}{\sqrt{2}} \Sigma^0 + \frac{1}{\sqrt{6}} \Lambda & n
%\\
%\Xi^- & \Xi^0 & - \frac{2}{\sqrt{6}} \Lambda
%\end{array}
%\right)
%\label{eq:baryons}
%\end{equation}
%
%
At lowest order in momentum, that we will keep in our study, the interaction
Lagrangian reads
%comes from the $\Gamma_{\mu}$ term in the covariant
%derivative
%and we find
%
\begin{equation}
L_1^{(B)} = < \bar{B} i \gamma^{\mu} \frac{1}{4 f^2}
[(\Phi \partial_{\mu} \Phi - \partial_{\mu} \Phi \Phi) B
- B (\Phi \partial_{\mu} \Phi - \partial_{\mu} \Phi \Phi)] > \ ,
\label{eq:lowest}
\end{equation}
where $\Phi$ and $B$ are the SU(3) matrices for the mesons and baryons,
respectively and the symbol $< >$ stands for the trace of the resulting SU(3)
matrix. The Lagrangian of Eq.~(\ref{eq:lowest})
leads to a common structure of the type
$\bar{u} \gamma^u (k_{\mu} + k'_{\mu}) u$ for the different channels, where
$u, \bar{u}$ are the Dirac spinors and $k, k'$ the momenta of the incoming
and outgoing mesons.

We take the $K^- p$ state and all those that couple to it within the chiral
scheme, namely $\bar{K}^0 n$, $\pi^0 \Lambda$, $\pi^0 \Sigma^0$,
$\pi^+ \Sigma^-$, $\pi^- \Sigma^+$, $\eta \Lambda$, $\eta
\Sigma^0$, $K^0\Xi^0$ and
$K^+\Xi^-$.  Hence we have a problem with ten coupled channels. 

The lowest-order amplitudes for these channels are easily evaluated from 
Eq.~(\ref{eq:lowest}) and are given by
\begin{equation}
V_{i j} = - C_{i j} \frac{1}{4 f^2} \bar{u} (p_i) \gamma^{\mu} u (p_j)
(k_{j\mu} + k_{i\mu})
\label{eq:ampl}
\end{equation}
where $p_j, p_i (k_j, k_i)$ are the initial, final momenta of the baryons
(mesons).
For low energies one can  write this amplitude as 
\begin{equation}
%       V_{i j} = - C_{i j} \frac{1}{4 f^2} (k^0 + k'^0) \, .
V_{i j} = - C_{i j} \frac{1}{4 f^2}(2\sqrt{s} - M_{Bi}-M_{Bj})
\left(\frac{M_{Bi}+E}{2M_{Bi}}\right)^{1/2} \left(\frac{M_{Bj}+E^{\prime}}{2M_{Bj}}
\right)^{1/2}\, ,
\label{eq:ampl2}
\end{equation}
and the matrix $C_{i j}$, which is symmetric, is given in \cite{angels}.

Note that the use of physical masses in Eq.~(\ref{eq:ampl2}) is
introducing effectively some contributions of higher orders in the
chiral counting. In the standard chiral approach one would be using the average
mass of the octets in the chiral limit and higher order Lagrangians involving
SU(3) breaking terms would generate the mass differences. By introducing the
physical masses one guarantees that the phase space for the reactions,
thresholds and unitarity in coupled channels are respected from the beginning.
We also use in our approach an average value for the pseudoscalar meson decay
constant, $f=1.15f_\pi$, as done in \cite{angels}. 
%Another source of symmetry
%breaking is
%introduced later on to fine tune the data by changing one of the $a_i$
%parameters.

We shall construct the amplitudes using the isospin formalism for
which we must use average masses for the $K$ $(K^0,K^+)$, $\bar{K}$ $(K^-,
\bar{K}^0$), $N$
$(p,n)$, $\pi$ ($\pi^+,\pi^0,\pi^-$), $\Sigma$
($\Sigma^+,\Sigma^0,\Sigma^-$) and $\Xi$ $(\Xi^-,\Xi^0)$  states.
The isospin states are given in \cite{angels}.

We have four $I = 0$ channels, $\bar{K} N, \pi \Sigma$, $\eta
\Lambda$ and $K \Xi$,
while there are five $I = 1$ channels,
$\bar{K} N, \pi \Sigma, \pi \Lambda, \eta \Sigma$ and $K \Xi$. 
The transition matrix elements in isospin formalism
read like Eq.~(\ref{eq:ampl2}) 
substituting the $C_{ij}$ coefficients by $D_{ij}$ for
$I=0$
and by $F_{ij}$ for  $I=1$, with the $D_{ij}$ and $F_{ij}$ coefficients
given in \cite{angels}.

In \cite{joseulf}, using the N/D method of \cite{nsd} for this particular case
it was proved that the scattering amplitude could be written by means of the
algebraic matrix equation 
\begin{equation}
T = [1 - V \, G]^{-1}\, V
\label{eq:bs1}
\end{equation}
with $V$ the matrix of Eq.~(\ref{eq:ampl2}) evaluated on shell, or
equivalently
\begin{equation}
T = V + V \, G \, T
\label{eq:bs2}
\end{equation}
with $G$ a diagonal matrix given by
\begin{eqnarray}
G_{l} &=& i \, \int \frac{d^4 q}{(2 \pi)^4} \, \frac{M_l}{E_l
(\vec{q}\,)} \,
\frac{1}{k^0 + p^0 - q^0 - E_l (\vec{q}\,) + i \epsilon} \,
\frac{1}{q^2 - m^2_l + i \epsilon} \nonumber \\
&=& \int^{q_{\rm max}} \, \frac{d^3 q}{(2 \pi)^3} \, \frac{1}{2
\omega_l(\vec{q}\,)}
\,
\frac{M_l}{E_l (\vec{q}\,)} \,
\frac{1}{p^0 + k^0 - \omega_l (\vec{q}\,) - E_l (\vec{q}\,) + i \epsilon}
\label{eq:gprop}
\end{eqnarray}
which depends on $p^0 + k^0 = \sqrt{s}$ and $q_{\rm max}$.

One can see that Eq. (\ref{eq:bs2}) is just the Bethe Salpeter equation 
but with the $V$ matrix factorized on shell, which allows one to extract the
scattering matrix $T$ trivially, as seen in Eq.~(\ref{eq:bs1}).

 The analytical expression for $G_l$ can be obtained from \cite{OOP97} using a
 cut off and from \cite{joseulf} using dimensional regularization, 
\begin{eqnarray} 
G_{l} &=& i 2 M_l \int \frac{d^4 q}{(2 \pi)^4} \,
\frac{1}{(P-q)^2 - M_l^2 + i \epsilon} \, \frac{1}{q^2 - m^2_l + i
\epsilon}  \nonumber \\ &=& \frac{2 M_l}{16 \pi^2} \left\{ a_l(\mu) + \ln
\frac{M_l^2}{\mu^2} + \frac{m_l^2-M_l^2 + s}{2s} \ln \frac{m_l^2}{M_l^2} +
\right. \nonumber \\ & &  \phantom{\frac{2 M}{16 \pi^2}} +
\frac{\bar{q}_l}{\sqrt{s}} 
\left[ 
\ln(s-(M_l^2-m_l^2)+2\bar{q}_l\sqrt{s})+
\ln(s+(M_l^2-m_l^2)+2\bar{q}_l\sqrt{s}) \right. \nonumber  \\
& & \left. \phantom{\frac{2 M}{16 \pi^2} +
\frac{\bar{q}_l}{\sqrt{s}}} 
\left. \hspace*{-0.3cm}- \ln(-s+(M_l^2-m_l^2)+2\bar{q}_l\sqrt{s})-
\ln(-s-(M_l^2-m_l^2)+2\bar{q}_l\sqrt{s}) \right]
\right\} \ ,
\label{eq:gpropdr}
\end{eqnarray}        
which has been rewritten
in a convenient way to show how the imaginary part of $G_l$ is
 generated and how one can go to the unphysical Riemann sheets in order to identify
 the poles. 
The dimensional regularization scheme is preferable
if one goes to higher energies where the on-shell momentum of the intermediate
states is not reasonably smaller than the cut off.

The coupled set of Eqs.~(\ref{eq:bs1}) 
were solved in \cite{angels} using a cut off momentum of 630
MeV in all channels. Changes in the cut off can be accommodated in terms
 of changes in $\mu$, the regularization scale in the dimensional
 regularization formula for  $G_l$, or in the subtraction constant
$a_l$. In
 order to obtain the same results as in \cite{angels} at low energies, we set
 $\mu$ equal to the cut off momentum of 630 MeV (in all channels) and then
find the values of the
 subtraction constants $a_l$ such as to have $G_l$ with the same value
with the
 dimensional regularization formula (Eq.~(\ref{eq:gpropdr})) and the cut
off formula (Eq.~(\ref{eq:gprop}))
at the $\bar{K} N$
 threshold. This determines the  values
\begin{equation} 
\begin{array}{lll} a_{{\bar K}N}=-1.84~~ &
a_{\pi\Sigma}=-2.00~~ & a_{\pi\Lambda}\,=-1.83 \\ a_{\eta
\Lambda}\,\,=-2.25~~ & a_{\eta\Sigma}=-2.38~~ & a_{K\Xi}=-2.52 \ . 
\end{array} 
\label{eq:coef}
\end{equation}
In this
way we guarantee that we obtain the same results at low energies as
in \cite{angels} and we find indeed that this is the case when
we repeat the calculation with the new $G_l$ of Eq.~(\ref{eq:gpropdr}).
Then we extend the results at
higher energies, looking for the eventual appearance of new
resonances. 
 
\begin{figure}[ht]
\begin{center}
\includegraphics[width=10cm]{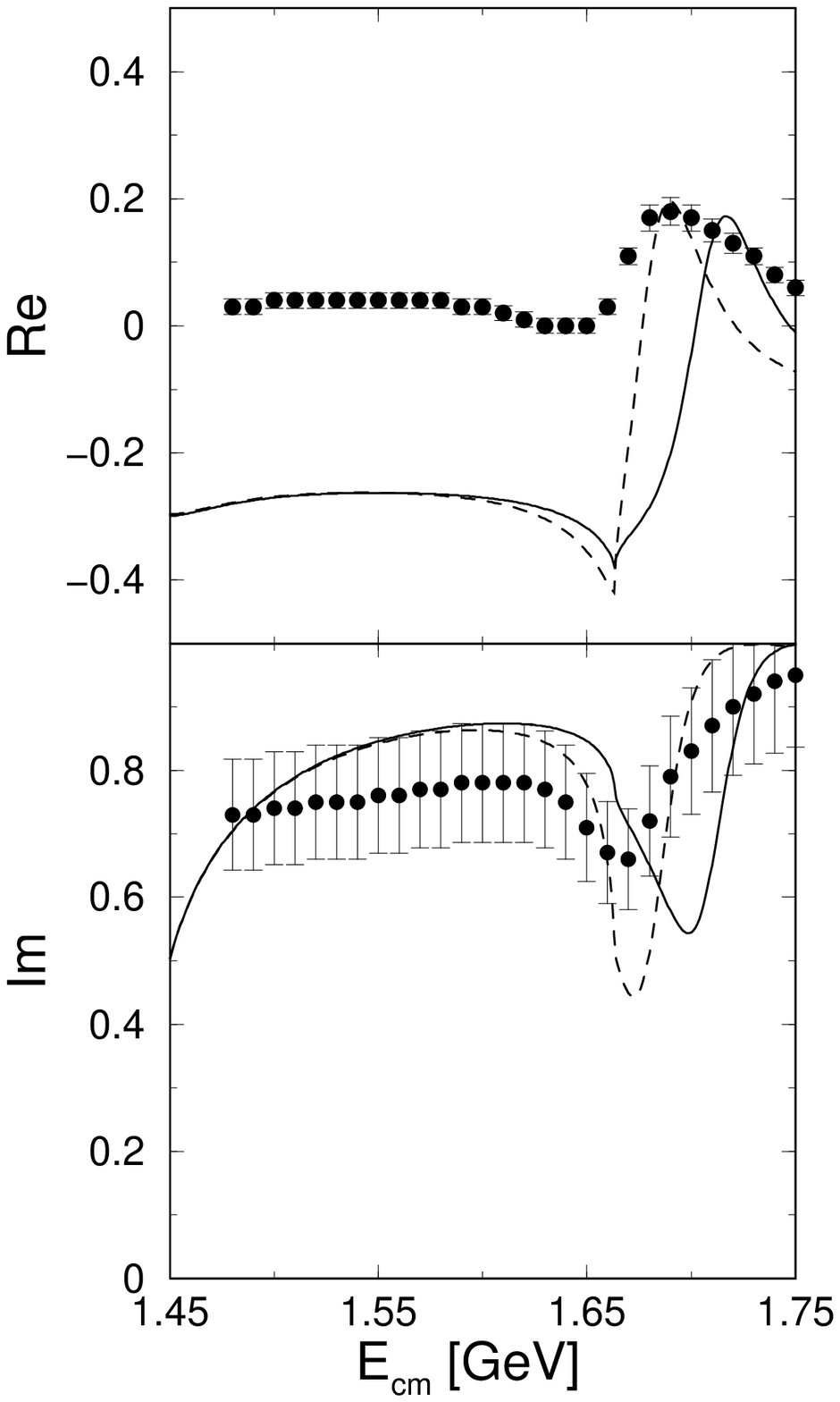}
\caption{$\bar{K}N \to \bar{K}N$ amplitude in the $I=0$ channel.
Data are taken from Ref.~\cite{gopal}.
}
\label{fig:KN0}
\end{center}
\end{figure}

\begin{figure}[ht]
\begin{center}
\includegraphics[width=10cm]{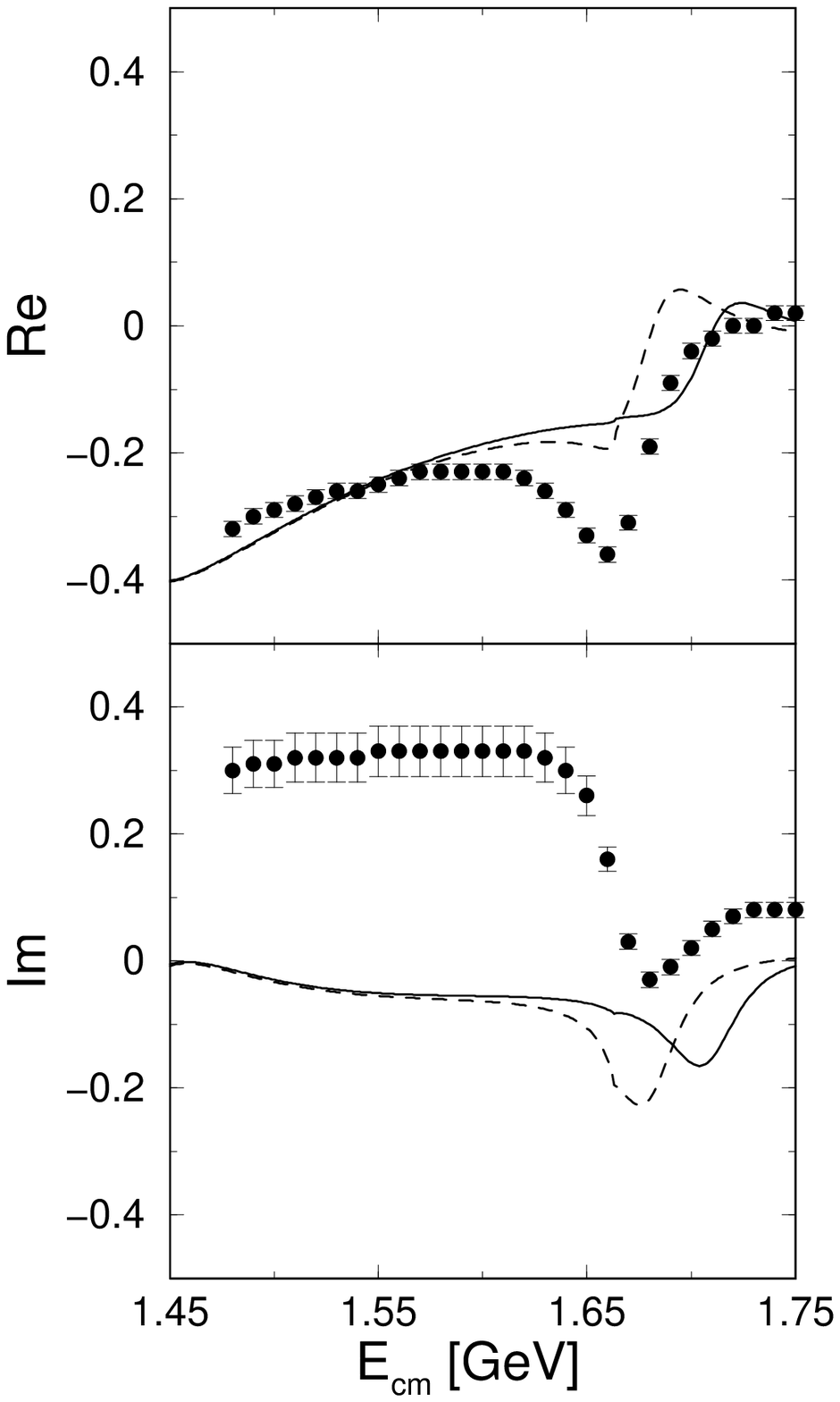}
\caption{$\bar{K}N \to \pi\Sigma$ amplitude in the $I=0$ channel.
Data are taken from Ref.~\cite{gopal}.
}
\label{fig:piS}
\end{center}
\end{figure}

The solid lines of Figs.~\ref{fig:KN0} and \ref{fig:piS} 
 show the results for the real and imaginary parts of
 the $I=0$ amplitudes for $\bar{K}N \to \bar{K}N$ and $\bar{K}N \to \pi
 \Sigma$, respectively.  Both channels clearly display the signal from
 the $\Lambda(1670)$ resonance, although large background contributions are 
 present in the amplitudes as well.
 
  The normalization of the amplitudes shown in Figs. \ref{fig:KN0} and 
\ref{fig:piS}  is different from
 the one of Eq.~(\ref{eq:bs1}).  
We shall call $T_M$ the plotted amplitudes and the
 relationship to our former amplitudes is given by
\begin{equation}
T_{M,ij}=-T_{ij}\frac{\sqrt{M_iM_j p_ip_j}}{4\pi \sqrt{s}}
\label{eq:tmat}
\end{equation}
 The normalization of $T_M$ is particularly suited to analyze the data in
 terms of the speed plot \cite{hohler}.  One has the amplitude written as 
\begin{equation}
T_{M,ij}(W)=T^{BG}_{M,ij}(W) -
\frac{x \,\, \Gamma/2\,\, {\rm e}^{i\phi} }{ W-M_R + i\Gamma/2 }
\label{eq:tres}
\end{equation}
 with $x=\sqrt{\Gamma_i\Gamma_j}/\Gamma$, where $\Gamma$, $\Gamma_i$,
$\Gamma_j$ are, respectively, the total width and
the partial decay widths of the resonance into the $i, j$ channels.
 
  The speed is defined as 
\begin{equation}
Sp_{ij}(W)=\left| \frac{dT_{M,ij}}{dW} \right| \simeq \frac{x 
 \Gamma/2 } {(W-M_R)^2 + \Gamma^2/4} \ ,  
\label{eq:speed}
\end{equation}
 where the second equality assumes that the background is smoothly dependent on
 the energy and does not contribute significantly to the derivative.
 
\begin{figure}[ht] 
\begin{center} 
\includegraphics[width=10cm]{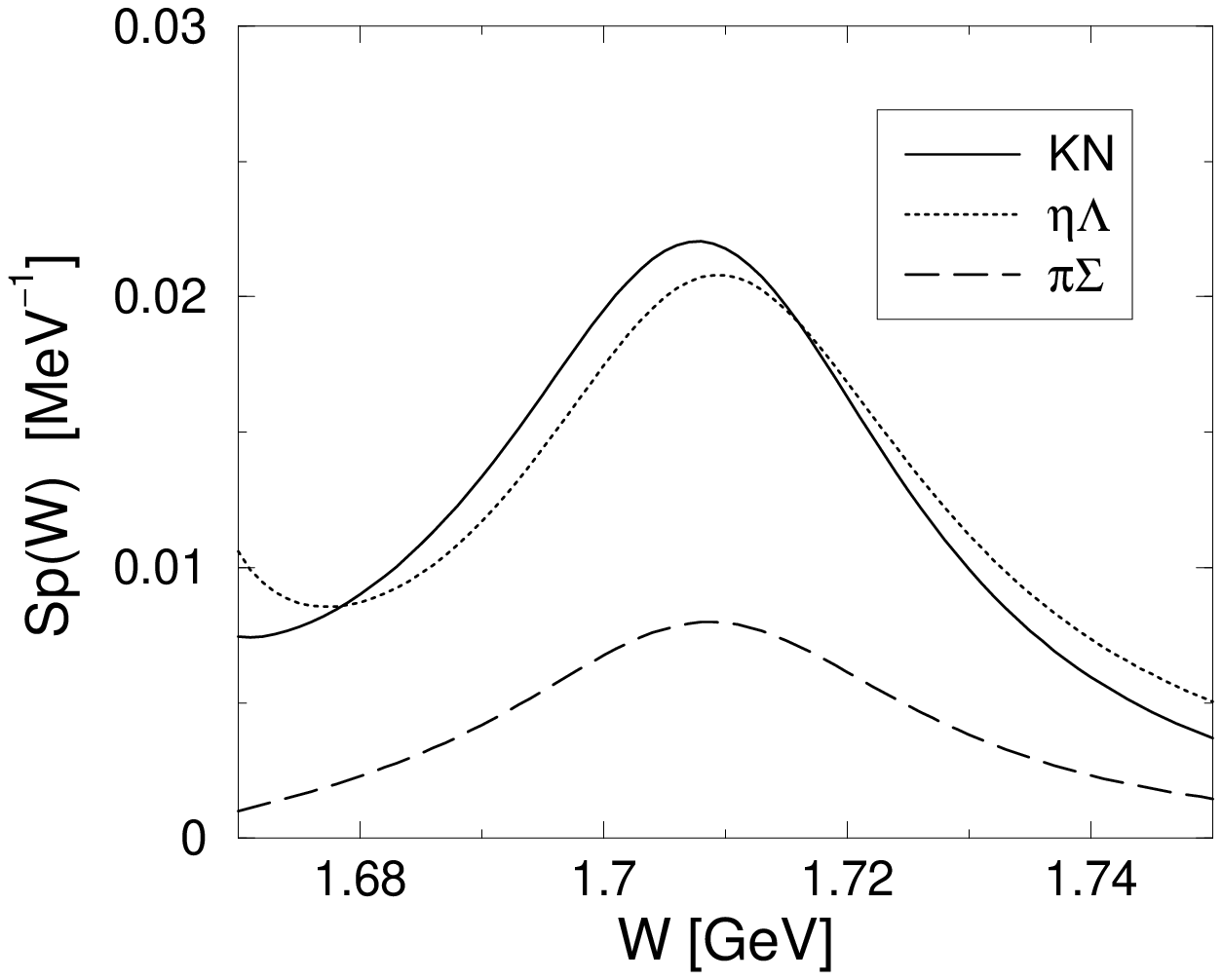}
\caption{Speed plot for the amplitudes $\bar{K}N \to \bar{K}N$ (solid
line), $\bar{K}N
\to \eta \Lambda$ (dotted line) and 
 $\bar{K}N \to \pi \Sigma$ (dashed line).
}
\label{fig:speed}
\end{center}
\end{figure}

In Fig.~\ref{fig:speed} we show the obtained speed
$Sp_{ij}(W)$ for different transitions,
$\bar{K}N \to
\bar{K}N$ (solid line), $\bar{K}N \to \eta \Lambda$ (dotted line) and
 $\bar{K}N \to \pi \Sigma$ (dashed line). 
As is evident from the plots in Fig.~\ref{fig:speed}, the background
induced by the already opened meson-baryon channels in the region of
interest is quite smooth, since an approximate Breit-Wigner shape
is obtained from the derivative of the amplitudes. 
On the other hand,
the resonance region we study does indeed lie above the two-pion threshold
for both $\Lambda$ and $\Sigma$ production. Such threshold openings could show
up in some form as a non-smooth background contribution. However, there is
no empirical evidence that any of the s-channel hyperon resonances under
investigation here couple strongly to the two-pion channel, hence, we do not
expect the Breit-Wigner shape of Fig.~\ref{fig:speed} to be modified by
the inclusion
of extra
inelastic channels.

The
study of the speed plots shown in Fig.~\ref{fig:speed} allows us to
obtain
 the energy $M_R=1708$ MeV, the total width $\Gamma=40$ MeV,
 and the branching ratios, $B_{\bar{K}N}=48$\%,
 $B_{\eta \Lambda}=45$\%, 
 and $B_{\pi \Sigma}=7$\%.
 Experimentally, one has $M_R= 1660 - 1680$ MeV, $ \Gamma=25-50$ MeV,
 $B_{\bar{K}N}=15-25$\%,
 $B_{\eta \Lambda}=15-35$\%, 
 and $B_{\pi \Sigma}=15-35$\% \cite{pdg}.
We seem to overestimate the $\bar{K}N$ and $\eta
\Lambda$ branching ratios and underestimate the $\pi \Sigma$ one.

Comparison of our
results with the experimental data in Figs.~\ref{fig:KN0} and \ref{fig:piS}
shows qualitative but not quantitative agreement. This is not surprising
since no parameters 
have been fitted to these data, but rather we have chosen the low-energy
parametrization of
 our theory which contained only one free parameter, the cut off to 
regularize the loops.  We now exploit the freedom that the
theory has by
 changing the parameters $ a_l$.  However, this must be accomplished in a
way
that does not ruin the very good agreement with the low-energy data, 
found in Ref.\cite{angels}.
We find that the results at low energies are very insensitive to 
 changes in the $a_{K\Xi}$ parameter, but they are sensitive to
changes in $a_{{\bar K}N}$, $a_{\pi \Sigma}$ and $a_{\eta\Lambda}$. On the
other hand, the position of the $\Lambda(1670)$ resonance is quite
sensitive to changes in the $a_{K \Xi}$ parameter and only moderately
sensitive to $a_{\bar{K}N}$, $a_{\pi \Sigma}$ and $a_{\eta\Lambda}$.
This allows us to fine tune the parameter $a_{K\Xi}$ (without changing the other
$a_{l}$ parameters) in order to better reproduce the position of the
resonance
found by experiment while maintaining the agreement found at low energies. 
Figs.~\ref{fig:KN0} and \ref{fig:piS}
also display the results using
$a_{K\Xi}=-2.67$ (dashed lines).  We see that a change of 6\% in this
parameter
moves the position of the resonance by $28$ MeV and it agrees better with
experiment.  The values of the resonance mass, width and branching ratios
obtained now are $M_R=1679$ MeV, the total width $\Gamma=40$ MeV,
 and the branching ratios $B_{\bar{K}N}=61$\%,
 $B_{\eta \Lambda}=30$\%, 
 and $B_{\pi \Sigma}=9$\%.

Comparison of the theoretical results for ${\bar K}N \to {\bar K}N$ with the
data shows agreement for
the imaginary part within errors, while our prediction for 
the real part below the resonance differs from the data by what appears
to be a large constant background term. This discrepancy needs to be
looked at with some perspective. The contribution of the real part to the cross
section  from the experimental data is negligible and is only 10 percent
in the theoretical case. On the other hand, our results around
$\sqrt{s}=1440$ MeV, the $\bar{K}N$ threshold, are in good agreement
with the data for the $K^-p$ and $K^-n$ scattering lengths, which would
suggest some discrepancy at low energies between the data shown in Fig.~\ref{fig:KN0}
\cite{gopal} and those of \cite{martin} and   \cite{iwasaki}.

In Fig.~\ref{fig:piS} we display the $I=0$ $\bar{K}N \to \pi\Sigma$
amplitude. The  
theoretical amplitude shows the resonance features with the same pattern
as the experiment, both for the real as for the imaginary parts. Yet there
is disagreement with the data in the imaginary part, again with an apparent
background missing for the theoretical prediction. Once again, the
discrepancy looks puzzling since up to $\sqrt{s}=1460$ MeV, and even
beyond where the s-wave is still dominant, the agreement of the present model
with the experimental cross sections for $K^-p \to \pi^-\Sigma^+, \pi^+
\Sigma^-, \pi^0 \Sigma^0$ is very good \cite{angels}.

We should note that the errors plotted in Figs.~\ref{fig:KN0} and \ref{fig:piS} 
correspond to the reasonable guesses of Ref.~\cite{penner}, but there are actual
deviations between the data of \cite{gopal}, \cite{langbein} and \cite{alston}.
Due to the large background in the experimental analysis, the
interference effects with the resonance are more apparent, leading to a
larger branching ratio to the $\pi\Sigma$ channel than the theory
predicts.

    We can also compare the results of the model with the recent
data on the $K^-p \to \eta\Lambda$ reaction \cite{nefkens}, which improve
on the older experiments \cite{old}. The shape of the results and the
position of the peak that we obtain agree well with the data for a 
parameter $a_{K\Xi}=-2.67$ but we
get a strength at the peak of $\sigma=2.7$ mb, about a factor of two larger
than the latest experimental value of
$\sigma=1.4$ mb. This reflects the fact that our predicted
 $\bar{K}N$ branching ratio overestimates the experimental value.

We have studied the reactions $K^- p \to K^+ \Xi^-$ and $K^- p \to K^0
\Xi^0$, which take place at energies beyond $\sqrt{s}=1.815$ GeV, hence above
the position of the $\Lambda(1670)$ resonance. Around a laboratory $K^-$
momentum of $1.6$ GeV/c, and using $a_{K\Xi}=-2.67$, our model predicts
a cross section of $0.17$ mb for the reaction
$K^- p \to K^+ \Xi^-$, which compares favourably with the experimental value
of $0.16-0.18$ mb \cite{griselin,debellefon}. For the reaction
$K^- p \to K^0 \Xi^0$ we find
a cross section of $0.24$ mb at $1.6$ GeV/c, which overestimates by
almost a factor of three the experimental value of $0.08-0.1$ mb
\cite{debellefon}. 
 
   In the  $I=1$ channel we find only rough agreement with the data in the 
 $\bar{K}N \to \bar{K}N$ and $\bar{K}N \to \pi\Sigma$ amplitudes, but,
 just like the data, we find no evidence of a resonance signal that would allow
 us to identify the $\Sigma(1620)$ resonance.  Clearly, the absence of 
 a signal even in some of the experimental amplitudes has lead to classifying the
$\Sigma(1620)$ as only a 2-star resonance.
 However, the absence of such a resonance would be  somewhat surprising since 
we expect to get an octet of meson-baryon
 resonances and so far only a singlet and the $I=0$ part of the octet
(eventually  mixed between themselves) have appeared.  Since we do not see
this state in the
 amplitudes at real energies we look for a pole in the
complex plane. We
 go directly to the second Riemann sheet, which we take in our case as the
one where the momenta of the channels which are open at energy $W$, with
${\rm Re}(z)=W$,  are taken negative in $G_l$. 
 
    Near the poles the amplitudes that we are analyzing behave as
\begin{equation}
T_{ij} \simeq \frac{g_i g_j}{z-z_R} ~~~~;~~~~T_{M,ij} \simeq 
\frac{x \Gamma/2 {\rm e}^{i\phi^\prime} }{z-z_R} \ .
\label{eq:tres2}
\end{equation}    
 Thus, the residues of the $T_{ij}$ matrix give the product of the coupling
 of the  resonance to the $i,j$  channels, while the residues of the
$T_{M,ij}$ 
 give one half  of the product of the two partial decay widths. The first
one of 
equations (\ref{eq:tres2}) determines the coupling of the  resonance
to
different final states, which are well defined  even
 if these states are closed in the decay of the resonance.
 
   The search of the poles  leads us, using $a_{K\Xi}=-2.52$, to the
values
$M_R=1708+i21$ MeV, 
$\Gamma=42$ MeV,
 $B_{\bar{K}N}=47$\%,
 $B_{\eta \Lambda}=47$\%, 
 and $B_{\pi \Sigma}=6$\%,
in remarkable agreement with the values obtained from the speed plot.
For  $a_{K\Xi}=-2.67$, we obtain
$M_R=1680 + i 20$ MeV, 
$\Gamma=40$ MeV,
 $B_{\bar{K}N}=54$\%,
 $B_{\eta \Lambda}=38$\%, 
 and $B_{\pi \Sigma}=8$\%.

 The couplings obtained for the $\Lambda(1670)$ resonance, using
 $a_{K\Xi}=-2.52$, are
\begin{equation}
\mid g_{{\bar K}N}\mid^2=0.51~~~~
\mid g_{\pi\Sigma}\mid^2=0.052~~~~
\mid g_{\eta\Lambda}\mid^2=1.0~~~~
\mid g_{K\Xi}\mid^2=11 \ ,
\label{eq:coup1}
\end{equation}
and, using $a_{K\Xi}=-2.67$, we obtain
\begin{equation}
\mid g_{{\bar K}N}\mid^2=0.61~~~~
\mid g_{\pi\Sigma}\mid^2=0.073~~~~
\mid g_{\eta\Lambda}\mid^2=1.1~~~~
\mid g_{K\Xi}\mid^2=12  \ .
\label{eq:coup2}
\end{equation}
It is also interesting to display the results of the complex plane search
for the $\Lambda(1405)$ resonance. We find
\begin{equation}
M_R=(1426 + i 16)~{\rm MeV}~~(\Gamma=32~{\rm MeV})   \nonumber
\end{equation}
\begin{equation}
\mid g_{{\bar K}N}\mid^2=7.4~~~~
\mid g_{\pi\Sigma}\mid^2=2.3~~~~
\mid g_{\eta\Lambda}\mid^2=2.0~~~~
\mid g_{K\Xi}\mid^2=0.12 \nonumber \ , \nonumber
\label{eq:coup3}
\end{equation}
with only the $\pi\Sigma$ channel open for the decay.

We have also performed a search in the $I=1$ channel and we indeed find
 a pole at 
\begin{equation}
M_R=(1579 + i 264)~{\rm MeV}~~(\Gamma\sim 528~{\rm MeV}) \ , \nonumber
\end{equation}
from the model with $a_{K\Xi}=-2.67$.
The couplings obtained are
\begin{equation}
\mid g_{{\bar K}N}\mid^2=2.6~~~~
\mid g_{\pi\Sigma}\mid^2=7.2~~~~
\mid g_{\pi\Lambda}\mid^2=4.2~~~~
\mid g_{\eta\Sigma}\mid^2=3.5~~~~
\mid g_{K\Xi}\mid^2=12 \ . \nonumber
\label{eq:coup4}
\end{equation}
 
 We find that the agreement with the PDG \cite{pdg} is quite good for the case
 of the $\Lambda(1405)$.  For the case of the $\Lambda(1670)$ we find a good 
agreement with
 the total width, but   the partial decay
widths to the $\bar{K}N $ and $\eta \Lambda$ channels is somewhat
overpredicted while the partial
 decay width to the $\pi \Sigma$ channel that we obtain is a bit too 
small.  For the case of the $\Sigma(1620)$
resonance we find a very large width which may be the reason why this state
does not provide a clearer signal in the scattering amplitudes.

   The analysis of the couplings is very interesting. In the case of the 
 $\Lambda(1405)$ state the coupling to the $\bar{K}N$ channel is found to be
 very large, while the coupling to the other channels is very small. This would
 allow us to identify this resonance as a quasibound $\bar{K}N$ state in the present
 approach.  Similarly, we find that the $\Lambda(1670)$ resonance has a large
 coupling to the $K\Xi$ channel and unusually small couplings to the other
 final states. This is responsible for the small width of the resonance in spite of
 the large phase space open for decay into the different channels. The large coupling
 to the $K\Xi$ channel allows identifying this state as a $K\Xi$ quasibound
 state in the present approach.  By contrast, the $\Sigma(1620)$ resonance
has couplings of normal size to all channels, and, given the large phase
space
 available, it has a sizable decay width into any of the channels and hence a 
 considerably larger total width.

    In summary,  we have demonstrated that the chiral approach 
 to the $\bar{K}N$ and the other coupled channels, which proved so successful at
 low energies, extrapolates smoothly to higher energies and provides the basic
 features of the  scattering amplitudes, generating the resonances which would
 complete the states of the  nonet of the $J^P=1/2^-$ excited states.  The
qualitative
 description of the data without adjusting any parameters is telling us that the
 basic information on the dynamics of these processes is contained in the chiral
 Lagrangians. There is still some freedom left with the chiral symmetry 
 breaking terms. In 
our formulation they would go into the $a_l$ subtraction constants, and the
use
of different decay constants for each meson, by means of which one could 
obtain a
better description of the data. 
However, before proceeding in this
direction, and eventually introduce further chiral symmetry breaking terms, 
it would be important to sort out the apparent 
discrepancies between different sets of data. 
The analysis of the poles and the couplings of
 the  resonances to the different channels lead us to identify
 the strong coupling of
 the $\Lambda(1405)$ resonance to the $\bar{K}N$ state and the large coupling 
 of the $\Lambda(1670)$ resonance to the $K\Xi$ state, allowing us to classify
 these resonances as quasibound states of $\bar{K}N$ and $K\Xi$, respectively.

\subsection*{Acknowledgments}

E. O. and C. B. wish to acknowledge the hospitality of the University
of Barcelona and A. R. and C. B. that of the University of Valencia, where part 
of this work was done. We would also like to acknowledge some useful
discussions with J.A. Oller and U.G. Meissner.  This work is also
partly supported by DGICYT contract numbers BFM2000-1326, PB98-1247,
by the EU TMR network Eurodaphne, contract no. ERBFMRX-CT98-0169, and
by the US-DOE grant DE-FG02-95ER-40907.

\end{document}